\documentclass{PoS}
\usepackage{graphicx, axodraw}
\usepackage{amsmath}
\def\dirac#1{#1\llap{/}}

\def\logo{\parbox{\textwidth}{\flushright LPT-ORSAY 09-28\\\flushleft\raisebox{-5mm}{\hbox{\noindent\includegraphics{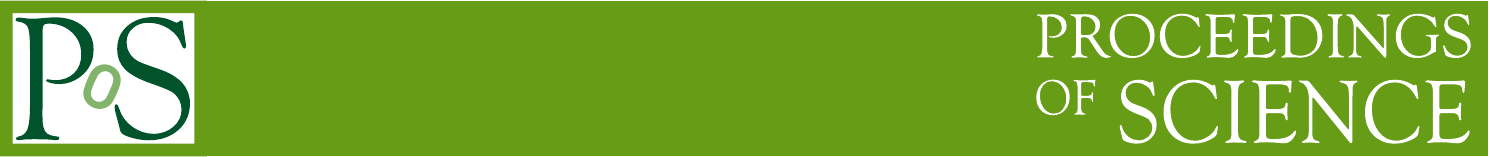}}}}}
\title{Renormalization of B-meson distribution amplitudes}

\ShortTitle{Renormalization of B-Meson distribution amplitudes}

\author{\speaker{N. Offen}\\
       Laboratoire de Physique Th\'eorique\\
CNRS/Univ. Paris-Sud 11 (UMR 8627),\\ F-91405 Orsay, France\\
        E-mail: \email{nils.offen@th.u-psud.fr}}

\author{S. Descotes-Genon \\
     Laboratoire de Physique Th\'eorique\\
CNRS/Univ. Paris-Sud 11 (UMR 8627),\\ F-91405 Orsay, France\\
        E-mail: \email{sebastien.descotes-genon@th.u-psud.fr}}

\abstract{We summarize a recent calculation of the evolution kernels of the two-particle $B$-meson distribution amplitudes $\phi_+$ and $\phi_-$ taking into account three-particle contributions. In addition to a few phenomenological comments, we give as a new result the evolution kernel of the combination of three-particle distribution amplitudes $\Psi_A-\Psi_V$ and confirm constraints on $\phi_+$ and $\phi_-$ derived from the light-quark equation of motion.}

\FullConference{International Workshop on Effective Field Theories: from the pion to the upsilon \\
		February 2-6 2009\\
		Valencia, Spain}

\begin{document}

\section{Introduction}
Exclusive decays of $B$-mesons provide important tools to test the Standard Model and to search for physics beyond it. Hadronic inputs encoding soft physics are not only form factors but also light-cone distribution amplitudes (LCDAs). In particular the $B$-meson LCDAs enter the parametrization of the hard-scattering part of hadronic matrix elements of bilocal current operators  where large momentum is tranferred to the soft spectator quark [1-10]. 
Impressive progress has been made in the calculation of the hard scattering amplitudes entering factorization theorems, see e.g. [11-15] for the $B\to P P$ case, 
but one limiting factor for the extraction of fundamental parameters is the uncertainty coming from the hadronic input. Recent years have seen several analyses concerning the renormalization properties \cite{Lange:2003ff,Bell:2008er,Braun:2003wx} and the shape of the $B$-meson LCDAs \cite{Grozin:1996pq, Braun:2003wx,  Khodjamirian:2005ea, Khodjamirian:2006st, Kawamura:2001jm, Lee:2005gza, Kawamura:2008vq}. Up to now these analyses were restricted to the two-particle case or to leading order with the exception of \cite{Kawamura:2008vq}. Here we present the results of \cite{DescotesGenon:2009hk} for the renormalization of the two-particle $B$-meson LCDAs taking into account mixing with three-parton LCDAs and in section (2.3) the results of a new calculation for the combination of three-particle LCDAs $\Psi_A-\Psi_V$ entering the equations of motion.

\section{One-loop calculation with three-parton external state}
The relevant two- and three-parton distribution amplitudes are defined as $B$ to vacuum matrixelements of a non-local heavy-to-light operator, which reads in the two-particle case \cite{Grozin:1996pq}
\begin{equation}
\langle 0|\bar{q}_\beta(z)[z,0] (h_v)_\alpha(0)|B(p)\rangle =
-i\frac{\hat{f}_B(\mu)}{4} 
  \left[(1+\dirac{v})\left(\tilde{\phi}_+(t)+\frac{\dirac{z}}{2t}[\tilde\phi_-(t)-\tilde\phi_+(t)] \right)\gamma_5\right]_{\alpha\beta}
\end{equation}
and in the three-particle case \cite{Kawamura:2001jm} (the most general decomposition without contraction with a light-like vector is given in \cite{Geyer:2005fb}):
\begin{eqnarray}
&&\langle 0|\bar{q}_\beta(z)[z,uz] gG_{\mu\nu}(uz)z^\nu[uz,0]
(h_v)_\alpha(0)|B(p)\rangle\\ \nonumber
&&\qquad\qquad =\frac{\hat{f}_B(\mu) M}{4}
  \Big[(1+\dirac{v})
  \Big[(v_\mu \dirac{z} -
    t\gamma_\mu)\left(\tilde\Psi_A(t,u)-\tilde\Psi_V(t,u)\right)
    -i\sigma_{\mu\nu}z^\nu \tilde\Psi_V(t,u)\\ \nonumber
&&\qquad\qquad\qquad\qquad\qquad
 - z_\mu \tilde{X}_A(t,u)
    +\frac{z_\mu \dirac{z}}{t} \tilde{Y}_A(t,u)
   \Big]\gamma_5\Big]_{\alpha\beta}.
\label{three-part-da}
\end{eqnarray}
We use light-like vectors $n_\pm$ so that every vector can be decomposed as
\begin{eqnarray}
q_\mu&=&(n_+\cdot q) \frac{n_{-,\mu}}{2} + (n_-\cdot q) \frac{n_{+,\mu}}{2} + q_{\perp\mu}
     =q_+ \frac{n_{-,\mu}}{2} + q_- \frac{n_{+,\mu}}{2} + q_{\perp\mu},\nonumber\\
n_+^2&=&n_-^2=0 \qquad n_+\cdot n_-=2 \qquad v=(n_++n_-)/2.
\end{eqnarray}
The computation of the renormalisation properties of the distribution
amplitudes requires us to consider matrix elements of the relevant operators
\begin{eqnarray}
O_+^H(\omega)&=&
  \frac{1}{2\pi}\int dt  e^{i\omega t }
    \langle 0 | \bar{q}(z) [z,0] \dirac{n}_+ \Gamma h_v(0) |H \rangle \\
O_-^H(\omega)&=&
  \frac{1}{2\pi}\int dt  e^{i\omega t }
    \langle 0 | \bar{q}(z) [z,0] \dirac{n}_- \Gamma h_v(0) |H \rangle \\
O_3^H(\omega,\xi)&=&
  \frac{1}{(2\pi)^2}\int dt  e^{i\omega t }\int du e^{i\xi ut } 
    \langle 0 | \bar{q}(z) [z,uz] g_s G_{\mu\nu}(uz) z^\nu [uz,0] \Gamma 
         h_v(0) |H \rangle 
\label{operators}
\end{eqnarray}
with $z$ parallel to $n_+$, i.e. $z_\mu=t  n_{+,\mu}$, 
$t =v\cdot z=z_-/2$ and the path-ordered exponential in 
the $n_+$ direction:
\begin{eqnarray} 
[z,0]&=&P\exp\left[i g_s \int_0^z dy_\mu A^\mu(y)\right]\\
 &=&1+ i g_s \int_0^1 d\alpha\ z_\mu A^\mu(\alpha z)
   -g_s^2 \int_0^1 d\alpha \int_0^\alpha d\beta z_\mu\ z_\nu\ 
           A^\mu(\alpha z)\ A^\nu(\beta z) + \ldots
\end{eqnarray}
The Fourier-transforms of the different distribution amplitudes are then defined via
\begin{equation}
\phi_\pm(\omega)= \frac{1}{2\pi}\int dt e^{i\omega t} \tilde\phi_\pm(t)\qquad
F(\omega,\xi)= \frac{1}{(2\pi)^2}\int dt\int du t e^{i(\omega+u\xi)t} \tilde{F}(t,u)
\end{equation}
where $F=\Psi_V,\Psi_A, X_A, Y_A$.
\begin{figure}
\begin{center}
\begin{picture}(120,60)(0,-10)
\SetWidth{1}
\Line(0,3)(60,58)
\Line(0,0)(60,55)
\ArrowLine(60,55)(120,0)
\Text(20,28)[]{$p$}
\LongArrow(110,15)(90,35)
\Text(110,28)[]{$k$}
\Gluon(60,0)(60,45){4}{4}
\LongArrow(70,0)(70,15)
\Text(90,5)[]{$\mu,\epsilon,a$}
\CCirc(60,55){10}{Black}{White}
\end{picture}

$\displaystyle
A : -g_s\frac{\epsilon_+}{q_+}
\left[\delta\left(\omega-k_+-q_+\right)-\delta\left(\omega-k_+\right)\right]
\bar{v}\dirac{n}_-\Gamma T^a u
$
\end{center}

\begin{center}
\begin{tabular}{ccc}
\begin{picture}(120,60)(0,-10)
\SetWidth{1}
\Line(0,3)(60,58)
\Line(0,0)(60,55)
\Line(60,55)(120,0)
\Gluon(60,0)(30,25){4}{4}
\CCirc(60,55){10}{Black}{White}
\end{picture}
&&
\begin{picture}(120,60)(0,-10)
\SetWidth{1}
\Line(0,3)(60,58)
\Line(0,0)(60,55)
\Line(60,55)(120,0)
\Gluon(60,0)(90,25){4}{4}
\CCirc(60,55){10}{Black}{White}
\end{picture}\\
$\displaystyle
B :- g_s\frac{v\cdot \epsilon}{v\cdot q}
\delta\left(\omega-k_+\right)
\ \bar{v}\dirac{n}_-\Gamma T^a u
$
&&
$\displaystyle
C : g_s\frac{1}{(k+q)^2}
\delta\left(\omega-k_+-q_+\right)
\ \bar{v}\dirac{\epsilon}(\dirac{k}+\dirac{q})\dirac{n}_-\Gamma T^a u
$
\end{tabular}
\end{center}
\caption{The three leading-order contributions to the matrix element of $O_\pm$ with a three-parton external state.}\label{fig:LO3part}
\end{figure}
Since the renormalization of the operators is independent of the infrared properties of the matrix-elements, we can choose an on-shell partonic external state consisting of a light quark, a heavy quark and a gluon in equation (\ref{operators}). The resulting leading-order diagrams are shown in figures \ref{fig:LO3part} and \ref{fig:LOO3}. Next-to-leading order (NLO) diagrams are obtained by adding a gluon or a quark loop (a ghost loop) in all possible places (for a complete list of diagrams, see \cite{DescotesGenon:2009hk}). Since the operators give rise to $\delta$-distributions in the $+$-component of the momenta, we chose to proceed via the theorem of residues. To be more explicit, we decomposed the loop momentum $l$ in light-cone components, picked up the poles in the $l_-$-integral and performed the $l_\perp$-integration in dimensional regularization with $D=2-2\epsilon$ dimensions. Additional $\frac{1}{\epsilon}$-poles arise through the $l_+$-integration for diagrams where a gluon is exchanged between the Wilson-line from the operator and the heavy-quark field. These are related to the cusp anomalous dimension, see e.g \cite{Korchemsky:1987wg, Korchemskaya:1992je}, stemming from the intersection of one light-like Wilson line from the path ordered exponential in the operator and one time-like Wilson line from the interaction of soft gluons with the heavy quark. The additional poles give rise to $\frac{1}{\epsilon^2}$-terms as well as Sudakov logarithms.\\
\begin{figure}
\begin{center}
\begin{tabular}{c}
\begin{picture}(120,75)(0,-10)
\SetWidth{1}
\Line(0,3)(60,58)
\Line(0,0)(60,55)
\ArrowLine(60,55)(120,0)
\Text(20,28)[]{$p$}
\LongArrow(110,15)(90,35)
\Text(110,28)[]{$k$}
\Gluon(60,0)(60,45){4}{4}
\LongArrow(70,0)(70,15)
\Text(90,5)[]{$\mu,\epsilon,a$}
\CCirc(60,55){10}{Black}{White}
\Text(60,55)[]{$3$}
\end{picture}\\
$\displaystyle
A_{3\mu} : ig_s(q_+ \epsilon_\mu - q_\mu \epsilon_+)\ \bar{v}\Gamma T^a u\delta(\omega-k_+)\delta(\xi-q_+)
$
\end{tabular}
\end{center}
\caption{Leading-order contribution to the matrix element of $O_{3\mu}$ with a three-parton external state.}\label{fig:LOO3}
\end{figure}
\subsection{$\phi_+$-case}
In the $\phi_+$-case there is no mixing from three-particle distribution amplitudes. $\gamma_{+,3}=0$ at order $\alpha_s$. We confirm the result for the anomalous-dimension matrix found in \cite{Lange:2003ff}
\begin{equation}
\gamma_+^{(1)}(\omega,\omega';\mu)\,=\,  \left(\Gamma^{(1)}_{\rm cusp}\log\frac\mu\omega+\gamma^{(1)} \right)\delta(\omega-\omega')
 -\Gamma^{(1)}_{\rm cusp}\omega
    \left(\frac{\theta(\omega'-\omega)}{\omega'(\omega'-\omega)}
            +\frac{\theta(\omega-\omega')}{\omega(\omega-\omega')}\right)_+
\label{gammapl}
\end{equation}
with
\begin{equation}
\Big[f(\omega,\omega')\Big]_+\,=\,f(\omega,\omega')-\delta(\omega-\omega')\int d\omega' f(\omega,\omega')\qquad\Gamma_{\rm cusp}^{(1)}\,=\,4\qquad \gamma^{(1)}\,=\,-2\nonumber
\end{equation}
\subsection{$\phi_-$-case}
The $\phi_-$ case is more involved. After coupling-constant and external leg renormalization there remains a genuine three-particle term. The renormalization group equation to order $\alpha_s$ can be written as
\begin{eqnarray}
\frac{\partial \phi_-(\omega;\mu)}{\partial \log\mu}&=&-\frac{\alpha_s(\mu)}{4\pi}\left(\int d\omega'\, \gamma^{(1)}_-(\omega,\omega';\mu)\phi_-(\omega';\mu)
\right.\nonumber\\
&&\left.
+\int d\omega'd\xi' \gamma_{-,3}^{(1)}(\omega,\omega',\xi';\mu) [\Psi_A-\Psi_V](\omega',\xi';\mu)\right)
\end{eqnarray}
where $\gamma_-^{(1)}$ is the result from \cite{Bell:2008er}
\begin{equation}
\gamma_-^{(1)}(\omega,\omega';\mu)\,=\,\gamma_+^{(1)}-\Gamma^{(1)}_{\rm cusp}\frac{\theta(\omega'-\omega)}{\omega'}
\end{equation}
and $\gamma_{-,3}^{(1)}$ from \cite{DescotesGenon:2009hk}
\begin{eqnarray}
\gamma_{-,3}^{(1)}(\omega,\omega',\xi')&=&4\left[\frac{\Theta(\omega)}{\omega'}\left\{(C_A-2C_F)\left[\frac{1}{\xi'^2}\frac{\omega-\xi'}{\omega'+\xi'-\omega}\Theta(\xi'-\omega)+\frac{\Theta(\omega'+\xi'-\omega)}{(\omega'+\xi')^2}\right]\right.\right.\nonumber\\
&&\qquad-\left.\left.C_A\left[\frac{\Theta(\omega'+\xi'-\omega)}{(\omega'+\xi')^2}-\frac{1}{\xi'^2}\left(\Theta(\omega-\omega')-\Theta(\omega-\omega'-\xi')\right)\right]\right\}\right]_+\nonumber\\
\label{gamma-3}
\end{eqnarray}
where we defined the $+$-distribution with three variables as
\begin{equation}
\Big[f(\omega,\omega',\xi')\Big]_+ =f(\omega,\omega',\xi')-\delta(\omega-\omega'-\xi')\int
d\omega f(\omega,\omega',\xi'')
\end{equation}
\subsection{$\Psi_A-\Psi_V$-renormalization and equation of motion constraints}
Here we report on an up to now unpublished calculation of the renormalization of the three-particle LCDAs $\Psi_A-\Psi_V$. We project on the relevant distribution amplitudes in equation (\ref{three-part-da}) using $\Gamma=\gamma_\perp^\mu\dirac{n}_+\dirac{n}_-\gamma_5$ (although a $\gamma^\mu$ instead of $\gamma_\perp^\mu$ yields the same result). The calculations go along the same lines as in the previous two cases, even though there is only one leading-order structure (shown in figure \ref{fig:LOO3}) and NLO diagrams must have one gluon attached to the vertex in order not to vanish trivially.  For convenience the result is splitted into $C_F$- and $C_A$-colour structures.
\begin{eqnarray}
\gamma_{3,3,C_A}^{(1)}(\omega,\xi,\omega',\xi')&=&2\left[\delta(\omega-\omega')\left\{\frac{\xi}{\xi'^2}\Theta(\xi'-\xi)-\left[\frac{\Theta(\xi-\xi')}{\xi-\xi'}\right]_+-\left[\frac{\xi}{\xi'}\frac{\Theta(\xi'-\xi)}{\xi'-\xi}\right]_+\right\}\right.\nonumber\\
&+&\delta(\xi-\xi')\left\{\left[\frac{\Theta(\omega-\omega')}{\omega-\omega'}\right]_++\left[\frac{\omega}{\omega'}\frac{\Theta(\omega'-\omega)}{\omega'-\omega}\right]_+\right\}+\delta(\omega+\xi-\omega'-\xi')\nonumber\\
&\times&\left\{\frac{1}{\xi'}\Theta(\omega-\omega')-\left[\frac{\Theta(\omega-\omega')}{\omega-\omega'}\right]_+-\left[\frac{\omega}{\omega'}\frac{\Theta(\omega'-\omega)}{\omega'-\omega}\right]_+\right\}\nonumber\\
&+&\delta(\omega+\xi-\omega'-\xi')\frac{1}{\xi'(\omega'+\xi')}\left\{\frac{\omega-\xi'}{\xi'}(\omega'+\xi'-\omega)\Theta(\omega-\omega')\right.\nonumber\\
&-&\left.\frac{\omega}{\omega'}(\omega'+2\xi'-\omega)\Theta(\omega'-\omega)\Theta(\omega)
+\frac{\omega}{\xi'}(\omega-\xi')\Theta(\xi'-\omega)\Theta(\omega)\right.\nonumber\\
&+&\left.\left.\frac{\omega-\xi'}{\omega'}(\omega'+\xi'-\omega)\Theta(\omega-\xi')\Theta(\xi)\right\}\right]
\label{CA33ev}
\end{eqnarray}
\begin{eqnarray}
\label{CF33ev}
\gamma_{3,3,C_F}^{(1)}(\omega,\xi,\omega',\xi';\mu)&=&\gamma_+^{(1)}(\omega,\omega';\mu)\delta(\xi-\xi')+\gamma_{R\, 3,3}^{(1)}(\omega,\xi,\omega',\xi')\nonumber\\
\gamma_{R\,3,3}^{(1)}(\omega,\xi,\omega',\xi')&=&4\delta(\omega+\xi-\omega'-\xi')\nonumber\\
&\times&\left[\frac{\xi^2}{\omega'}\frac{\Theta(\omega'-\xi)}{(\omega+\xi)^2}\Theta(\xi)+\frac{\omega}{\xi'}\frac{\Theta(\xi-\omega')}{\omega+\xi}\Theta(\omega)\left(\frac{\xi}{\omega+\xi}-\frac{\omega-\xi'}{\xi'}\right)\right]\nonumber\\
\end{eqnarray}
with $\gamma_+^{(1)}$ the same as in equation (\ref{gammapl}) and $\gamma_{3,3}^{(1)}$ defined as in (2.11) with obvious changes.
Part of this calculation, namely the light-quark-gluon part, has been calculated in a different context and a different scheme, e.g. in \cite{Bukhvostov:1985rn,Balitsky:1987bk}.\\
In \cite{Kawamura:2001jm} two equations from the light- and heavy-quark equations of motion were derived
\begin{equation}\label{eq:i}
\omega \phi_-'(\omega;\mu)+\phi_+(\omega;\mu)\,=\,I(\omega;\mu),\qquad(\omega-2\bar\Lambda)\phi_+(\omega;\mu)+\omega\phi_-(\omega;\mu)\,=\,J(\omega;\mu),
\end{equation} 
where $I(J)(\omega;\mu)$ are integro-differential expressions involving the three-particle LCDAs $\Psi_A-\Psi_V$ ($\Psi_A+X_A$ and $\Psi_V$) respectively. While the second equation was shown not to hold beyond leading order in \cite{Bell:2008er, Kawamura:2008vq} we have checked that the first one is valid once renormalization is taken into account. Taking the derivative of the first equation with respect to $\log\mu$, inserting
\begin{equation}
I(\omega;\mu)\,=\,2\frac{d}{d\omega}\int_0^\omega d\rho\int_{\omega-\rho}^\infty \frac{d\xi}{\xi}\frac{\partial}{\partial\xi} \left[\Psi_A(\rho,\xi;\mu)-\Psi_V(\rho,\xi;\mu)\right]
\end{equation}
and using the relation from \cite{Bell:2008er}
\begin{equation}
-\omega\frac{d}{d\omega}\int_0^\eta \frac{d\omega'}{\eta}\gamma_-^{(1)}(\omega,\omega';\mu)=\gamma_+^{(1)}(\omega,\eta;\mu)
\end{equation}
one arrives at the following equation
\begin{eqnarray}
&&\omega\frac{d}{d\omega}\int d\omega' d\xi' \gamma_{-,3}^{(1)}(\omega,\omega',\xi')\left(\Psi_A(\omega',\xi';\mu)-\Psi_V(\omega',\xi';\mu)\right)\nonumber\\
&+&2\int d\omega'\gamma_+^{(1)}(\omega,\omega';\mu)\frac{d}{d\omega'}\int_0^{\omega'}d\rho\int_{\omega'-\rho}^\infty \frac{d\xi}{\xi}\frac{\partial}{\partial\xi}\left(\Psi_A(\rho,\xi;\mu)-\Psi_V(\rho,\xi;\mu)\right)\nonumber\\
&=&2\int d\omega'd\xi'\frac{d}{d\omega}\int_0^\omega d\rho\int_{\omega-\rho}^\infty \frac{d\xi}{\xi}\frac{\partial}{\partial \xi}\gamma_{3,3}^{(1)}(\rho,\xi,\omega'\xi';\mu)\left(\Psi_A(\omega',\xi';\mu)-\Psi_V(\omega',\xi';\mu)\right),\nonumber\\
\end{eqnarray}
which can be proven to hold at order $\alpha_s$ by simple insertion of the respective evolution kernels (\ref{gammapl}), (\ref{gamma-3}), (\ref{CA33ev}), (\ref{CF33ev}). This non-trivial outcome gives us further confidence concerning the renormalization group properties of the LCDAs.
\section{Conclusions}
The presence of $\delta(\omega-\omega')\log(\mu/\omega)$ in the renormalization matrices gives rise to a radiative tail falling off like $(\log \omega)/\omega$ for large $\omega$. Therefore non-negative moments of the LCDAs are not well defined and have to be considered with an ultraviolett cut-off.\cite{Lange:2003ff, Bell:2008er,Lee:2005gza,Kawamura:2008vq}
\begin{equation}
\langle \omega^N\rangle_\pm(\mu)\,=\,\int_0^{\Lambda_{UV}} d\omega\,\omega^N\,\phi_\pm(\omega;\mu)
\end{equation}
For $\phi_-$ it is interesting to examine the limit 
\begin{equation}
\lim_{\Lambda_{UV}\to \infty}\int_0^{\Lambda_{UV}} d\omega\, \omega^N\, z_{-,3}^{(1)}(\omega,\omega',\xi')=0\qquad N=0,\,1,\quad  z_{-,3}^{(1)}=\frac{1}{2\epsilon}\gamma_{-,3}^{(1)},
\end{equation}
which is relevant for the calculation of the three-particle contributions to the moments:
\begin{eqnarray}
\int_0^{\Lambda_{UV}}d\omega\, \omega^N\,\phi_-(\omega;\mu) &=& 1+\frac{\alpha_s}{4\pi}\left(\int d\omega'\,\phi_-(\omega')\,\int_0^{\Lambda_{UV}}d\omega\, \omega^N\,z^{(1)}_-(\omega,\omega';\mu)\right.\\
&&-\left.\int d\omega'd\xi'(2-D)[\Psi_A-\Psi_V](\omega',\xi')\int_0^{\Lambda_{UV}}d\omega\,\omega^N\,z^{(1)}_{-,3}(\omega,\omega',\xi')\right)\nonumber
\end{eqnarray}
Therefore as stated in \cite{Bell:2008er} three-particle distribution amplitudes give only subleading contribution to the first two moments ($N=0,\,1$) and we have explicitly checked that this statement cannot be extended to higher moments $(N\geq 2)$.\\

The next step consists in using the renormalization properties as a guide to go beyond the existing models derived from a leading-order sum-rule calculation resulting in $\Psi_A=\Psi_V$ \cite{Khodjamirian:2006st} and to analyze their influence on $\phi_-$. Finally, for practical calculations involving three-particle contributions, one would need the evolution kernels for the all relevant LCDAs, which will be the subject of a future work. 
\section*{Acknowledgments}

Work supported in part by EU Contract No. MRTN-CT-2006-035482, \lq\lq FLAVIAnet'' and by the ANR contract \lq\lq DIAM'' ANR-07-JCJC-0031.

\end{document}